\documentclass{aa}
\usepackage{graphicx}
\usepackage{txfonts}
\usepackage{natbib}
\bibpunct{(}{)}{;}{a}{}{,}

\def \kms{km s$^{-1}$}

\begin{document}

\title{Resolving distance ambiguities towards 6.7 GHz methanol masers}
\author{J. D. Pandian\inst{1} \and E. Momjian\inst{2}\thanks{Now at NRAO, P.O. Box O, Socorro, NM 87801, U.S.A.} \and P. F. Goldsmith\inst{3}}
\institute{Max-Planck-Institut f\"{u}r Radioastronomie, Auf dem H\"{u}gel 69, 53121 Bonn, Germany\\
\email{jpandian@mpifr-bonn.mpg.de}
\and
Arecibo Observatory, HC3 Box 53995, Arecibo, PR 00612, U.S.A.\\
\email{emomjian@naic.edu}
\and
Jet Propulsion Laboratory, California Institute of Technology, Pasadena, CA 91109, U.S.A.\\
\email{Paul.F.Goldsmith@jpl.nasa.gov}}

\abstract
{Distances to most star forming regions are determined using kinematics, through the assumption that the observed radial velocity arises from the motion of the source with respect to the Sun resulting from the differential rotation of Galaxy. The primary challenge associated with the application of this technique in the inner Galaxy is the kinematic distance ambiguity.}
{In this work, we aim to resolve the kinematic distance ambiguity towards a sample of 6.7 GHz methanol masers, which are signposts of the early stages of massive star formation.}
{We measured 21 cm \ion{H}{i} absorption spectra using the Very Large Array in C and CnB configurations. A comparison of the maximum velocity of \ion{H}{i} absorption with the source velocity and tangent point velocity was used to resolve the kinematic distance ambiguity.}
{We resolved the distance ambiguity towards 41 sources. Distance determinations that are in conflict with previous measurements are discussed. The NE2001 spiral arm model is broadly consistent with the locations of the star forming complexes. We find that the use of vertical scale height arguments to resolve the distance ambiguity can lead to erroneous classifications for a significant fraction of sources.}
{}


\maketitle

\section{Introduction}
Measuring distances towards massive star forming regions is important both for characterizing the physical properties of these regions, and for studying the spiral structure of the Galaxy. Trigonometric parallax is the most direct and model independent technique for measuring distances to astronomical objects. However, it is difficult to measure trigonometric parallax towards a large sample of massive star forming regions due to their large distances. The popular method for deriving the distance to a source arbitrarily located in the disk of the Milky Way is to assume that the measured radial velocity of the source arises from its differential rotation in the Galaxy. Then, by using a model for the rotation of the Galaxy (e.g. \citealt{clem85}; C85 hereafter), one can obtain an estimate of the kinematic distance to the source. One of the main problems associated with kinematic distances is that the distance solution is double valued in the inner Galaxy. Thus, a given radial velocity for a source in the inner Galaxy corresponds to two distances -- a near distance and a far distance. The two solutions are equidistant about the tangent point, which is the location where the line of sight is a tangent to the assumed circular orbit of a source about the Galactic center. This problem is called the kinematic distance ambiguity (KDA) and is the primary obstacle towards a straightforward application of kinematic distances to study Galactic structure.

One of the techniques to resolve the KDA is to measure an absorption spectrum towards the source of interest. Since only material in front of the target can give rise to absorption, one can use the velocities of the absorption features to distinguish between the near and far distance (see Fig. 1 of \citealt{kolp03} for a schematic). Thus, if absorption features are seen at velocities between that of the source and of the tangent point, the source has to be at the far distance since the clouds producing absorption at these velocities will be located between the near and far distance solution points (and will consequently be further from the Earth than the near distance point). For the technique to work, the tracer used to measure the absorption has to be abundant in the Galaxy. The two popular tracers used for this purpose are \ion{H}{i} at 21 cm \citep{kuch94, fish03, kolp03}, and formaldehyde (CH$_2$O) at 6 cm \citep{wils72, aray02, wats03, sewi04}.

Each of the above tracers has its advantages and disadvantages. Formaldehyde is less widespread than \ion{H}{i}, and this can lead to an incorrect resolution of the KDA. One way this can come about is simply that there is a low formaldehyde abundance in clouds between the near and far distance points.  The resulting non--detection evidently leads to incorrect classification of a far distance source to be at the near distance. The low excitation temperature of the 6 cm formaldehyde transition means that no discrete background source is necessary, as formaldehyde absorption lines can be detected against the Cosmic Microwave Background (CMB). In contrast, \ion{H}{i} is much more abundant and widely distributed in the interstellar medium, so that a cloud lacking HI is a rare occurrence. Consequently, any KDA resolution that is obtained in this way is likely to be correct.  It is true that absorption spectra from single dish telescope observations are difficult to interpret due to emission from gas spatially extended beyond the boundaries of the background continuum source (e.g. \citealt{dick90}). This confusion can be mitigated through interferometer observations, as discussed in Section 3. Another potential is that young \ion{H}{ii} regions may lack 21 cm continuum emission against which to measure \ion{H}{i} absorption. In some cases, where the emission is optically thick, the continuum may be more easily detectable at 6 cm compared to 21 cm, but at least for observations of \ion{H}{i}, lack of sufficient continuum will be evident.  Thus, the consequence is simply a lack of data from such sources.

Masing transitions of molecules such as methanol, OH and water are frequently associated with early phases of massive star formation. Methanol masers at 6.7 GHz, the brightest of Class II methanol masers, are unique in that they appear to be exclusively associated with massive star formation (e.g. \citealt{mini03, pand07b}). Thus, methanol masers are potentially useful to further our understanding of Galactic spiral structure. This is bolstered by the recent measurement of the distance to 12.2 GHz methanol masers (which are the second brightest Class II methanol masers) in W3OH, and the Perseus spiral arm \citep{xu06}. It is however, not possible to apply the absorption spectrum technique to resolve the KDA towards a majority of methanol masers in the inner Galaxy. This is because the sources immediately associated with the masers often do not have detectable radio continuum (especially at long wavelengths) due to the young age of the massive young stellar objects (MYSOs). For example, only about 10\% of the 6.7 GHz methanol masers detected in the blind Arecibo survey have detectable continuum at 21 cm \citep{pand07b}.  However, it is well known that massive stars form in clustered environments. Thus, while the number of methanol masers directly associated with radio continuum may be small, it is likely that a larger fraction will be in a cluster hosting a more evolved MYSO that has associated radio continuum emission. Here, we report on the results of \ion{H}{i} observations with the Very Large Array (VLA) to resolve the KDA towards a sample of 6.7 GHz methanol masers that either have direct association with 21 cm continuum, or are located in a cluster hosting a 21 cm continuum source.

\section{Sample Selection}
The methanol maser sample was selected from the General Catalog of methanol masers \citep{pest05}. We first restricted ourselves to methanol masers in the first Galactic quadrant with Galactic longitude $l > 10\degr$. At low Galactic longitudes, a large fraction of sources in the line of sight have a significant component of their circular motion (around the Galactic center) oriented tangential to the Sun. Consequently, for these sources, radial peculiar motions give rise to large errors in the kinematic distance. Sources near the tangent point do not have this problem, but the difference between the near and far distance is not as large compared to sources at larger Galactic longitudes. Furthermore, these sources are at low declination where the sensitivity of the VLA is poorer. 

We searched for point sources in the NRAO VLA Sky Survey (NVSS; \citealt{cond98}) within $\sim 2\arcmin$ of each methanol maser. We placed a cut-off of 30 mJy for the flux of the continuum source in the interest of observing time required to obtain an absorption spectrum with good signal to noise ratio. We then looked for radio recombination line (RRL) associations with the continuum sources (using the catalog of \citealt{lock89}; L89 hereafter) to verify their nature as \ion{H}{ii} regions. The velocities of the recombination lines were also used to check the association of the \ion{H}{ii} region with the cluster hosting the methanol maser. Sources with no RRL information in L89 were observed with the National Radio Astronomy Observatory's\footnote{The National Radio Astronomy Observatory is a facility of the National Science Foundation operated under cooperative agreement by Associated Universities, Inc.} Green Bank Telescope (GBT) for this purpose. Sources with no RRLs were discarded as they are likely to be continuum sources other than \ion{H}{ii} regions that are projected against our target masers. These selection criteria gave a sample of 64 methanol masers that we observed with the VLA.

\section{Observations}
\subsection{GBT Observations and Data Reduction}
The observations were made in June 2005 using the C-Band receiver of GBT. The GBT spectrometer was used in 16 intermediate frequency (IF) mode to cover 8 frequencies in two polarizations. The IFs were centered on the hydrogen recombination lines H103$\alpha$ to H110$\alpha$. The frequencies adopted for the transitions were 5931.545 MHz, 5762.881 MHz, 5600.551 MHz, 5444.261 MHz, 5293.733 MHz, 5148.703 MHz, 5008.923 MHz and 4874.158 MHz respectively for the H103$\alpha$ to H110$\alpha$ lines. Each band had a bandwidth of 12.5 MHz with 4096 channels, giving a total velocity coverage of $\sim$ 650 \kms~with $\sim$ 0.2 \kms~resolution. The observations were made in frequency switched mode with a 3 MHz frequency throw. The integration time per source was typically 15 minutes, with shorter integrations (5--10 minutes) being used for sources with strong continuum. The system temperature was usually around 25 K. A pointing and focus check was done at the beginning of each observation. The flux scale was set using noise diodes in the receiver. Since the intent of the observations was to detect RRLs and their velocities, we did not do any additional flux calibration using a calibrator. We estimate the flux scale to be accurate to around 10\%.

The data were reduced using GBTIDL. Frequency switched observations yield two sets of difference spectra for each IF. Each difference spectrum was baseline subtracted using a 3rd order polynomial. The baseline subtracted difference spectra of the IFs that were not affected by radio frequency interference were then summed together, taking into account the different velocity resolutions at the different frequencies. The resulting difference spectra were then folded to give the final spectrum of the source. This technique gives spectra with fairly good baselines that have a signal to noise ratio as much as a factor of 4 higher than what would be achieved using a standard position switched observation of a single recombination line. A Gaussian fit was performed on the detected RRLs to determine their velocity with respect to the local standard of rest (LSR).

\subsection{VLA Observations and Data Reduction}
A total of 63 different fields (covering 64 sources) was observed with the VLA between June and October 2005. Sources at declinations below --15\degr~were observed in the CnB configuration, while the other sources were observed in the C configuration. The quasar 3C 286 was used as a primary flux calibrator. The correlator was configured to have a bandwidth of 1.5625 MHz with 256 channels in single IF mode. The calibrators were observed at +250 \kms~and --180 \kms~(with respect to the rest frequency of \ion{H}{i} in the LSR frame of reference, assumed to be 1420.40575 MHz) to avoid contamination from Galactic \ion{H}{i}, while the targets were observed at a velocity of +50 \kms. The total integration time per field was 10--25 minutes depending on the NVSS flux of the continuum source.

The data were reduced using the Astronomical Image Processing System (AIPS) of NRAO employing standard procedures. The initial calibration was done using the ``channel 0'' data, the solutions from which were copied to the line data. For each source, we summed line free channels to create continuum datasets. We then CLEANed and self-calibrated the continuum data, and copied the resulting solutions to the line data. We were concerned about CLEANing the line data since \citet{kolp03} reported that the procedure introduced spurious absorption features from the emission features that are not resolved by the interferometer, leading them to extract spectra from the dirty image cubes. While the latter procedure works for strong continuum sources that are not affected by the sidelobes of other sources in the field, it fails for the weaker sources in our sample.

Hence, we did an extensive test of CLEANing the line data. We used the strong sources as a reference to check for any introduction of spurious absorption features. We found that as long as the CLEAN is restricted to $\sim 2\sigma$ of the noise in a channel image, the procedure works without any artifacts being introduced. We also found that using boxes to restrict the area that is CLEANed did not work well, as it introduced \ion{H}{i} emission from outside the CLEAN boxes (which are centered around the continuum source) into the targets. Although the emission features lie outside the CLEAN boxes, they enter the CLEAN boxes through the sidelobes of the dirty beam. The use of CLEAN boxes would thus force the task to clean the sidelobe emission instead of the peak which lies outside the CLEAN boxes. The exception is when one is dealing with fields that have both weak and strong point sources that are embedded in a large extended structure. In these cases, a full field CLEAN did not reproduce the weak point sources, and introducing boxes to restrict the area CLEANed gave much better results. We also found that removing the continuum prior to the CLEAN (using the task UVLSF) yielded better results compared to CLEANing the fields with no continuum subtraction. All the spectra presented in this paper are derived from CLEANed image cubes. The CLEANed data were then divided by the continuum image to derive optical depth information. 

\section{KDA Resolution and Distance Calculation}

To resolve the KDA, we used the approach followed by \citet{kolp03}, who performed a simulation to explore the relation between the quality of KDA resolution and the tangent point velocity, $v_t$, source velocity, $v_s$, and the highest velocity of a \ion{H}{i} absorption feature, $v_a$, taking into account random velocities, streaming motions and cloud-cloud dispersions. They found that the KDA can be resolved with good confidence unless $(v_t-v_s) < (v_t-v_a) + 15$ \kms~{\it and} $v_t - v_a < 20$ \kms~(these are sources with velocities close to the tangent point velocity). Further, within the ambiguous region, the line $(v_t-v_a) = (v_t-v_s)/2$ separated the near distance and far distance sources at the confidence level of 50\% (Fig. 2 of \citealt{kolp03}). We use the same criteria, and assign sources to the near or far distance with the quality grade of A and B depending on whether the source lies outside or within the ambiguous region, respectively.

To calculate distances, we use two approaches. The first is the classical approach, employing the source velocity, $v_s$, to calculate the distance using the C85 rotation curve with $(R_0,~\Theta_0) = (8.5,~220)$. We prefer to use the C85 rotation curve as all our sources are in the first quadrant, and the C85 curve takes into account peculiar motions in this region, such as the dip in velocity near a Galactocentric radius of 3 kpc \citep{burt78}. The errors in the distance are calculated using the distances corresponding to $v_s \pm 10$ \kms~to allow for a the velocity of the region differing from that arising from pure circular rotation. 

However, parallax measurements of massive star forming regions show a consistent bias in that kinematic distances over-predict the true distance to the source \citep{reid08}. Using VLBI measurements of the proper motion of Sagittarius A$^*$, the value of $\Theta_0/R_0$ has been measured to be 29.5 \kms~kpc$^{-1}$ with 0.3\% accuracy, assuming that Sagittarius A$^*$ is stationary \citep{reid04}. This translates to a $\Theta_0$ of 250.8 \kms~for an $R_0$ of 8.5 kpc. The use of these new parameters still yields significant systematic peculiar motions for the young massive star forming regions as seen in the frame of reference that is rotating with the Galaxy. Doing a global fit to reduce the peculiar motions, \citet{reid08} find that massive star forming regions are born rotating $\sim 13$ \kms~slower than the Galactic rotation curve, and that the motion of the Sun towards the LSR is consistent with that derived by \citet{dahn98} from the Hipparcos data.

Hence, in an attempt to make kinematic distances closer to the parallax distance, we used the following approach: we first calculated the heliocentric velocity from the current LSR model (which assumes a solar motion of 20 \kms~towards ($\alpha$, $\delta$) = (19$^h$, 30\degr) in 1900 coordinates). We then recalculated an LSR velocity using the \citet{dahn98} model with $(U_0,~V_0,~W_0) = $ (10.0, 5.25, 7.17). Using the new velocity (typically less than the old velocity by $\sim 10$ \kms in the first and second Galactic quadrants), we calculated the distance using the C85 curve with $(R_0,~\Theta_0)$ = (8.5 kpc, $250.8-13$ \kms). To calculate uncertainties, instead of using the traditional approach of calculating kinematic distances corresponding to $v_s \pm 10$, we used a $\pm 10$ \kms~variation to the component of the source velocity that is radial to the Galactic center. This is because the peculiar motions of massive star forming regions, in addition to being $\sim 13$ \kms~slower to Galactic rotation, are preferentially oriented in the direction of the Galactic Center, as is seen for example in W3OH \citep{xu06}. We also included a minimum uncertainty of $\pm 5$ \kms~to the source velocity to take into account random motions, since the above prescription will not yield any uncertainty to the LSR source velocity when the source is near the tangent point. 

The difference between the distances calculated using the two methods is typically small ($\sim 0.2$ kpc), but can be larger than 0.5 kpc, especially at velocities near the tangent point velocity, and in the second Galactic quadrant.

\section{Results}
The continuum sources associated with (either directly or through a cluster) 36 methanol masers were detected with adequate signal to noise ratio to resolve the KDA. The remaining 28 sources were typically too resolved to extract spectra with good enough signal to noise ratio. Treating multiple recombination line velocities (towards the same line of sight) as separate sources, we have a total of 41 sources. The properties of these sources are listed in Table 1. Column (1) contains the source names. Sources which have two recombination line velocities in the same line of sight are suffixed A and B for distinction. Column (2) lists the NVSS source that is associated with the methanol maser. Column (3) shows the velocity extent of the maser emission, and column (4) shows the systemic velocity, $v_s$ derived from the radio recombination lines. Column (5) lists the KDA resolution specifying whether the source is at the near distance (N), far distance (F), tangent point (T), or whether the KDA is unresolved (?). Column (6) shows the distance $d$, calculated from the conventional method using the C85 rotation curve, while column (7) shows $d'$, which is the modified kinematic distance calculated using the methodology described in section 4. Column (8) indicates the quality of the KDA resolution (A or B), and column (9) lists the rms noise in the 21 cm continuum image. Errors for $d$ are not listed when $v_s+10$ exceeds the tangent point velocity, $v_t$, since such velocities are forbidden, and cannot be converted to a meaningful distance. 

Figure 1 shows the continuum images, \ion{H}{i} optical depth, and GBT recombination line data (where applicable) for our sample. The optical depth spectra also indicate the source velocity, $v_s$, and the tangent point velocity, $v_t$, as short and long dashed lines, respectively.

\begin{table*}
\begin{minipage}[t]{\textwidth}
\caption{6.7 GHz Methanol Maser Properties}
\centering
\renewcommand{\footnoterule}{}
\begin{tabular}{l c c c c c c c c}
\hline \hline
Source & NVSS & Maser $(v_{min},~v_{max})$ & $v_s$ & KDA & $d$ & $d'$ & Quality & $\sigma$ \\
Name & Source & (\kms) & (\kms) & Resolution & (kpc) & (kpc) & Assignment & mJy beam$^{-1}$ \\
(1) & (2) & (3) & (4) & (5) & (6) & (7) & (8) & (9) \\
\hline
10.32--0.15\footnote{Distance from \citet{blum01}} & 180900-200456 & (4, 17) & 11.0 & N & $3.4 \pm 0.3$ & $3.4 \pm 0.3$ & & 8.0 \\
10.47+0.02 & 180835--195219 & (58, 77) & 70.1 & F & $10.6 \pm +0.4$ & $10.8 \pm 0.3$ & A & 5.3 \\
10.62--0.38$^a$ & 181028--195548 & (--10, 7) & 0.4 & N & $3.4 \pm 0.3$ & $3.4 \pm 0.3$ & & 1.5 \\
10.95+0.02 & 180942--192629 & (23, 25) & 18.5 & F & $14.2^{+1.3}_{-1.0}$ & $14.3 \pm 1.2$ & A & 1.3 \\
11.03+0.06 & 180939--192116 & (15, 21) & 12.6 & F & $15.0 \pm 1.3$ & $15.0 \pm 1.3$ & A & 1.3 \\
11.93--0.61 & 181401--185324 & (30, 44) & 39.3 & N & $4.2 \pm 0.8$ & $4.0 \pm 0.7$ & A & 2.2 \\
11.99--0.27 & 181245--184039 & (60, 61) & 58.2 & F & $11.3 \pm 0.6$ & $11.5 \pm 0.6$ & A & 1.4 \\
12.20--0.10A & 181242--182417 & (2, 15) & 27.0 & F & $13.5 \pm 1.0$ & $13.6^{+1.0}_{-0.7}$ & A & 1.3 \\
12.20--0.10B & 181242--182417 & & 53.6 & F & $11.6 \pm 0.6$ & $11.8 \pm 0.6$ & A & 1.3 \\
12.68--0.18 & 181355--180124 & (50, 62) & 57.4 & N & $5.1 \pm 0.6$ & $4.9 \pm 0.6$ & A & 5.5 \\
12.79--0.19 & 181414--175543 & (31, 41) & 35.8 & N & $3.8^{+0.6}_{-0.9}$ & $3.6 \pm 0.7$ & A & 2.5 \\
12.90--0.26 & 181444--175223 & (35, 47) & 28.6 & F & $13.4 \pm 1.0$ & $13.6 \pm 0.9$ & A & 2.5 \\
13.18+0.06 & 181405--172832 & (48, 49) & 54.1 & N & $4.8 \pm 0.6$ & $4.6 \pm 0.5$ & A & 2.5 \\
14.09+0.10 & 181545--163858 & (4, 17) & 16.3 & F & $14.7 \pm 1.1$ & $14.8 \pm 0.9$ & A & 1.5 \\
18.06+0.08A\footnote{Source has ambiguity between two recombination lines} & 182335--130920 &  & 16.4 & F? & $14.7 \pm 0.9$ & $14.8 \pm 0.8$ & A & 2.0 \\
18.06+0.08B$^b$ & 182335--130920 & (45, 57) & 60.1 & F? & $11.7 \pm 0.5$ & $11.9 \pm 0.4$ & A & 2.0 \\
18.46--0.00 & 182436--125106 & (47, 50) & 56.5 & F & $11.9 \pm 0.5$ & $12.1 \pm 0.4$ & A & 2.2 \\
19.48+0.15 & 182604--115234 & (20, 27) & 19.8 & F & $14.4 \pm 0.8$ & $14.5 \pm 0.7$ & A & 2.2 \\
19.61--0.13 & 182716--115322 & (52, 54) & 58.6 & F & $11.8 \pm 0.5$ & $12.0 \pm 0.4$ & A & 1.7 \\
23.43--0.18 & 183438--083043 & (94, 113) & 104.5 & F & 9.2 & $9.6 \pm 0.3$ & B & 3.3 \\
24.50--0.03 & 183606--073106 & (108, 116) & 109.2 & F & 8.6 & $9.3^{+0.3}_{-0.5}$ & A & 1.5 \\
24.67--0.14 & 183652--072453 & (111, 114) & 111.4 & N & 7.2 & $6.4 \pm 0.5$ & B & 2.4 \\
24.78+0.08 & 183610--071124 & (106, 115) & 108.6 & N & 6.8 & $6.1 \pm 0.4$ & B & 1.6 \\
25.70+0.04A & 183804--062523 &  & 53.3 & F & $11.8 \pm 0.6$ & $12.0 \pm 0.4$ & A & 1.0 \\
25.70+0.04B & ? & (89, 101) & 102.0 & ? & $\ldots$ & $\ldots$ &  & 1.0 \\
27.28+0.15 & 183610--071124 & (34, 37) & 36.3 & F & $12.7 \pm 0.7$ & $12.9 \pm 0.5$ & A & 0.8 \\
28.30--0.38A\footnote{The continuum source associated with the A and B sources may be reversed} & 184415--041757 &  & 44.8 & F & $12.1 \pm 0.6$ & $12.3 \pm 0.5$ & A & 0.8 \\
28.30--0.38B$^c$ & 184422--041746 & (79, 93) & 75.6 & F & $10.3 \pm 0.6$ & $10.6 \pm 0.3$ & A & 0.8 \\
30.70--0.07 & 184735--020143 & (85, 90) & 101.0 & F & 8.5 & $9.0 \pm 0.4$ & A & 4.4 \\
30.76--0.05 & 184738--015731 & (88, 95) & 91.6 & F & 8.5 & $9.0 \pm 0.4$ & B & 4.4 \\
30.82--0.05 & 184746--015451 & (91, 110) & 105.7 & F & 8.5 & $9.0 \pm 0.4$ & B & 4.4 \\
32.97+0.04 & 185124+000412 & (89, 93) & 93.2 & F & 8.3 & $8.7 \pm 0.4$ & B & 2.0 \\
33.13--0.09 & 185207+000819 & (71, 81) & 93.8 & F & 8.2 & $8.7 \pm 0.4$ & B & 1.3 \\
33.40+0.01A & 185220+002552 &  & 72.3 & F & $9.8 \pm 0.7$ & $10.0 \pm 0.4$ & A & 1.3 \\
33.40+0.01B & 185216+002531? & (96, 107) & 105.4 & T & 7.1 & 7.1 & A & 1.3 \\
41.10--0.22 & 190711+070956 & (63, 64) & 59.4 & F & $9.1 \pm 0.8$ & $9.4 \pm 0.4$ & A & 1.9 \\
43.87-0.77 & 191426+092236 & (47, 53) & 55.0 & F & $8.8 \pm 0.8$ & $9.1 \pm 0.5$ & A & 2.7 \\
45.07+0.13 & 191320+105054 & (57, 60) & 63.0 & F & 7.7 & $8.0 \pm 0.5$ & B & 1.4 \\
48.90-0.27\footnote{Distance from \citet{genz81}} & 192214+140319 & (63, 73) & 66.5 & ? & $7.0 \pm 1.5$ & $7.0 \pm 1.5$ &  & 1.4 \\
50.29+0.69 & 192127+154406 & (26, 33) & 25.9 & F & $9.3 \pm 0.7$ & $9.3 \pm 0.3$ & A & 1.3 \\
81.87+0.78\footnote{$d'$ for this source is the distance to the tangent point} & 203837+423759 & (0, 13) & 1.7 & F & 2.0 & 1.2 & B & 2.9 \\
\hline
\end{tabular}
\begin{flushleft}Column (1) contains the source names. Sources which have two recombination line velocities in the same line of sight are suffixed A and B for distinction. Column (2) lists the NVSS source that is associated with the methanol maser. Column (3) shows the velocity extent of the maser emission, and column (4) shows the systemic velocity, $v_s$ derived from the radio recombination line. Column (5) lists the KDA resolution specifying whether the source is at the near distance (N), far distance (F), tangent point (T), or whether the KDA is unresolved (?). Column (6) shows the distance $d$, calculated from the conventional method using the C85 rotation curve, while column (7) shows $d'$, which is the modified kinematic distance calculated using the methodology described in section 4. Column (8) indicates the quality of the KDA resolution (A or B), and column (9) lists the rms noise in the 21 cm continuum image. Errors for $d$ are not listed when $v_s+10$ exceeds the tangent point velocity, $v_t$, since such velocities are forbidden, and cannot be converted to a meaningful distance.
\end{flushleft}
\end{minipage}
\end{table*}

\begin{figure*}[h]
\centering
\includegraphics[width=17cm]{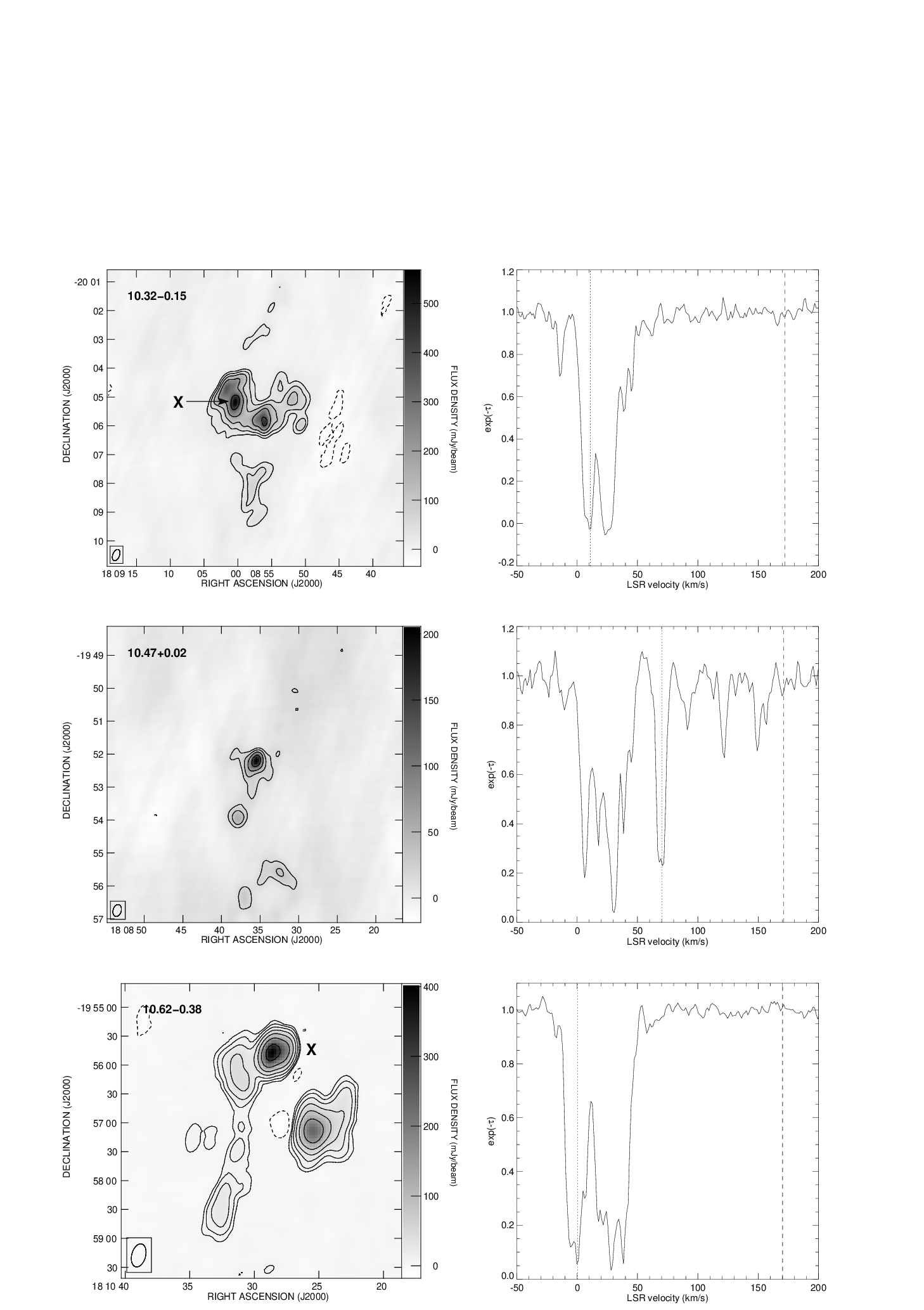}
\caption{Images of the 21 cm continuum, \ion{H}{i} optical depth spectra and GBT recombination line spectra (where applicable) of the sources in our study. The left panel shows the continuum image with contours at --2 and 2 times the rms noise listed in column (9) of Table 1, and at factors of 2 thereafter. For sources with multiple velocity components, the component from which the \ion{H}{i} spectrum is derived is marked with an `X'. The right panel shows the line to continuum ratio, or $\exp(-\tau)$. The source velocity, $v_s$, and the tangent point velocity, $v_t$, are indicated by short and long dashed lines, respectively. For sources with GBT recombination line data, the top right panel shows the \ion{H}{i} spectrum, while the bottom right panel shows the recombination line spectrum in emission.}\label{fig1}
\end{figure*}

\setcounter{figure}{0}
\begin{figure*}
\centering
\includegraphics[width=17cm]{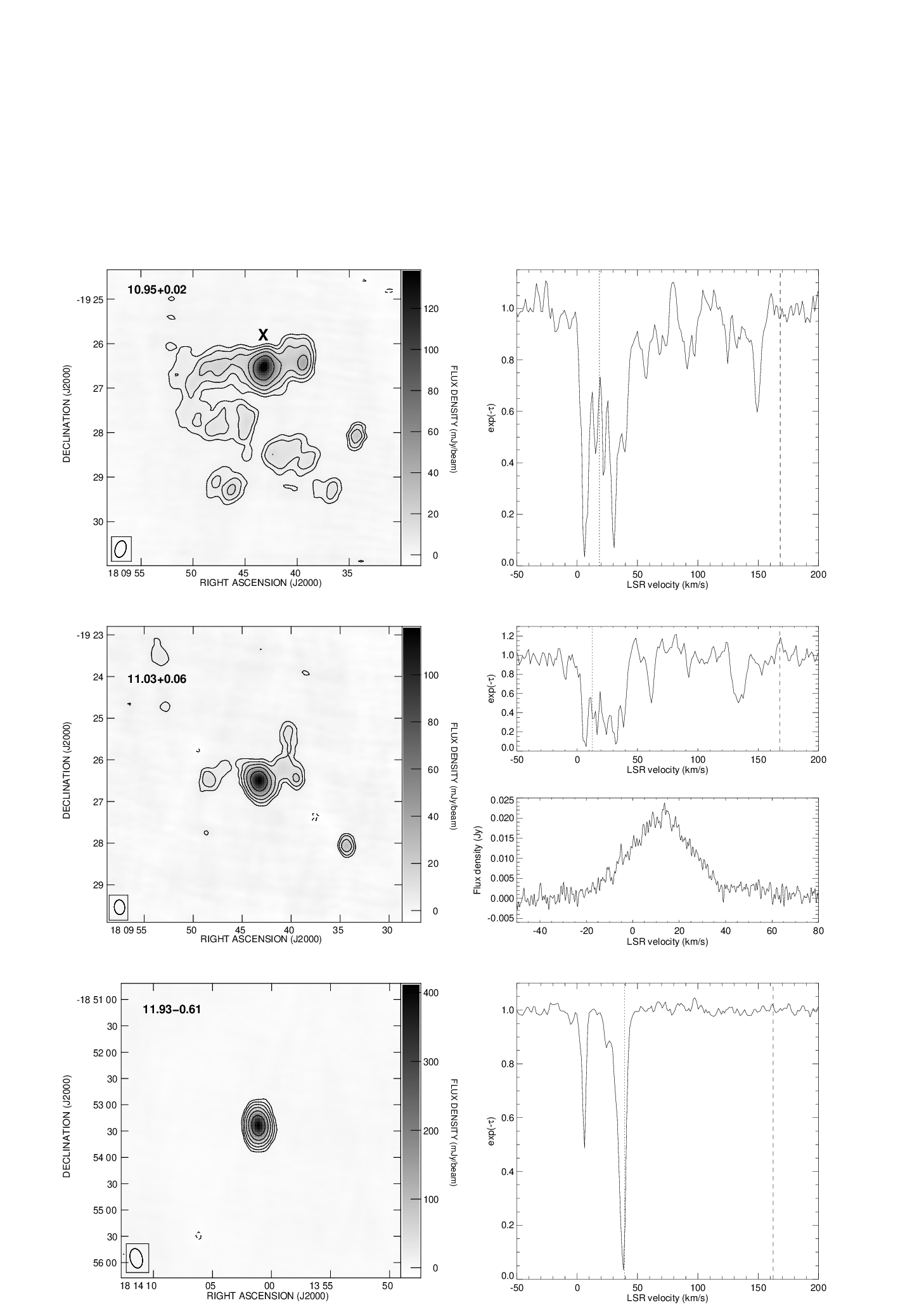}
\caption{Continued.}
\end{figure*}

\setcounter{figure}{0}
\begin{figure*}
\centering
\includegraphics[width=17cm]{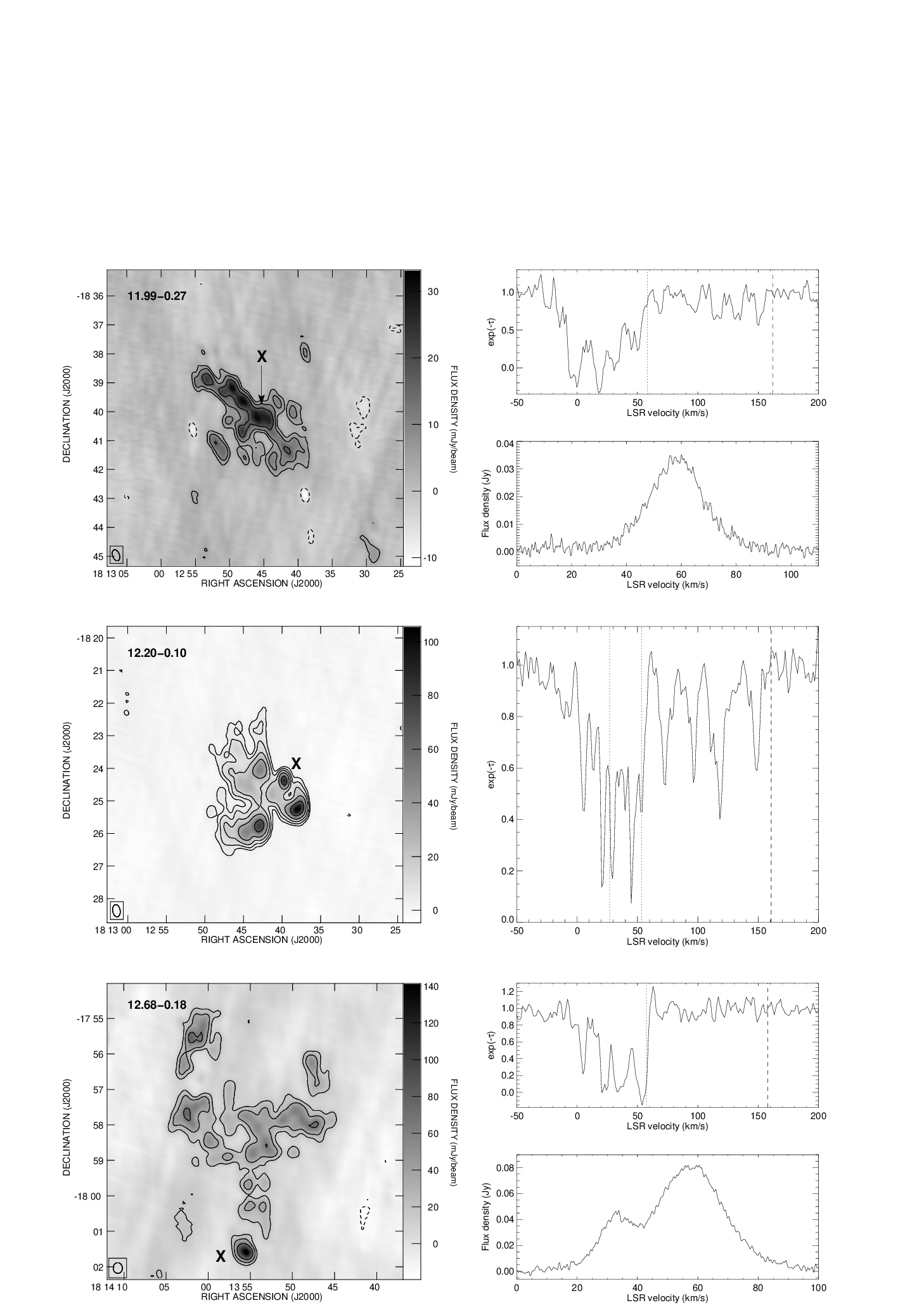}
\caption{Continued.}
\end{figure*}

\setcounter{figure}{0}
\begin{figure*}
\centering
\includegraphics[width=17cm]{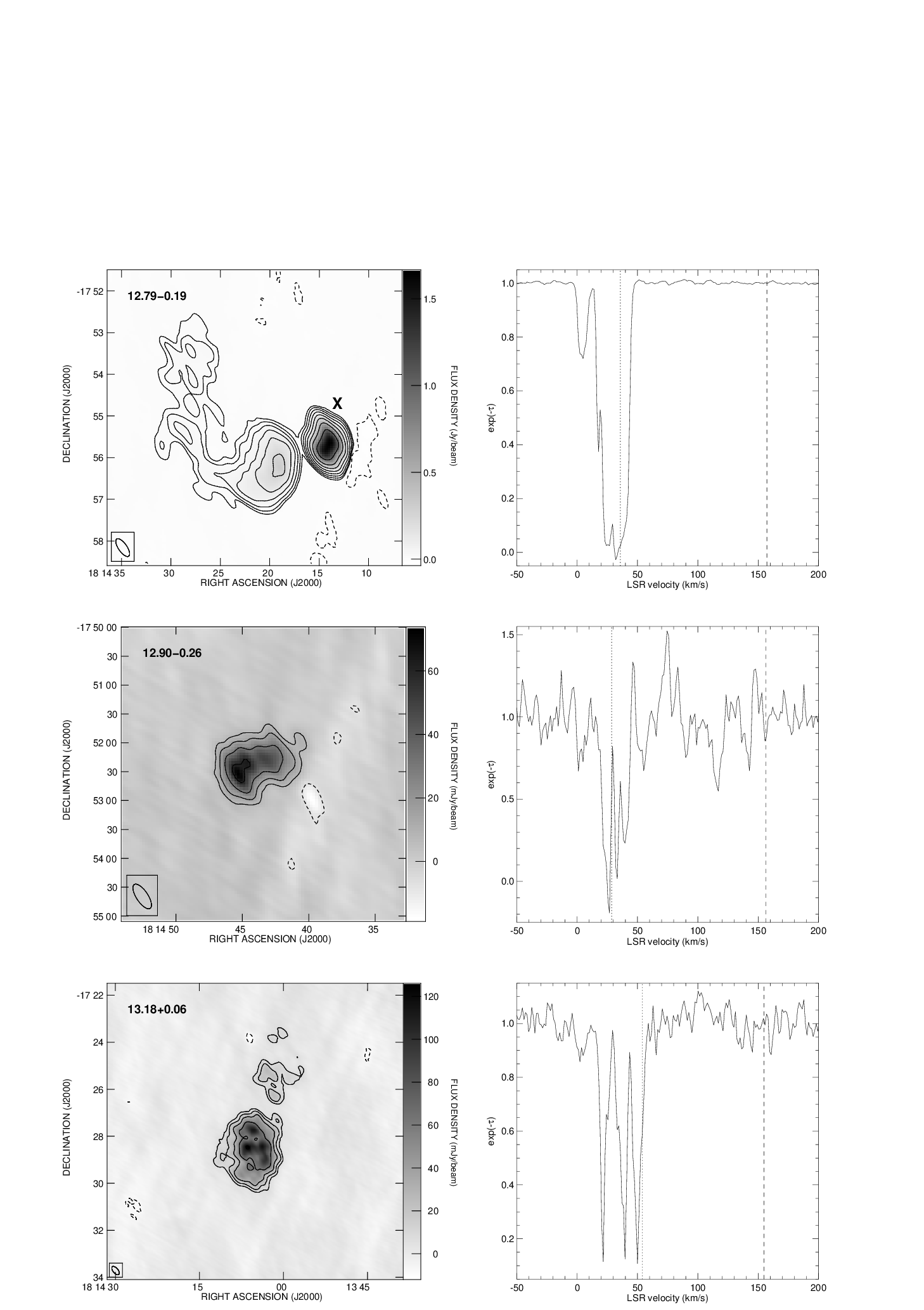}
\caption{Continued.}
\end{figure*}

\setcounter{figure}{0}
\begin{figure*}
\centering
\includegraphics[width=17cm]{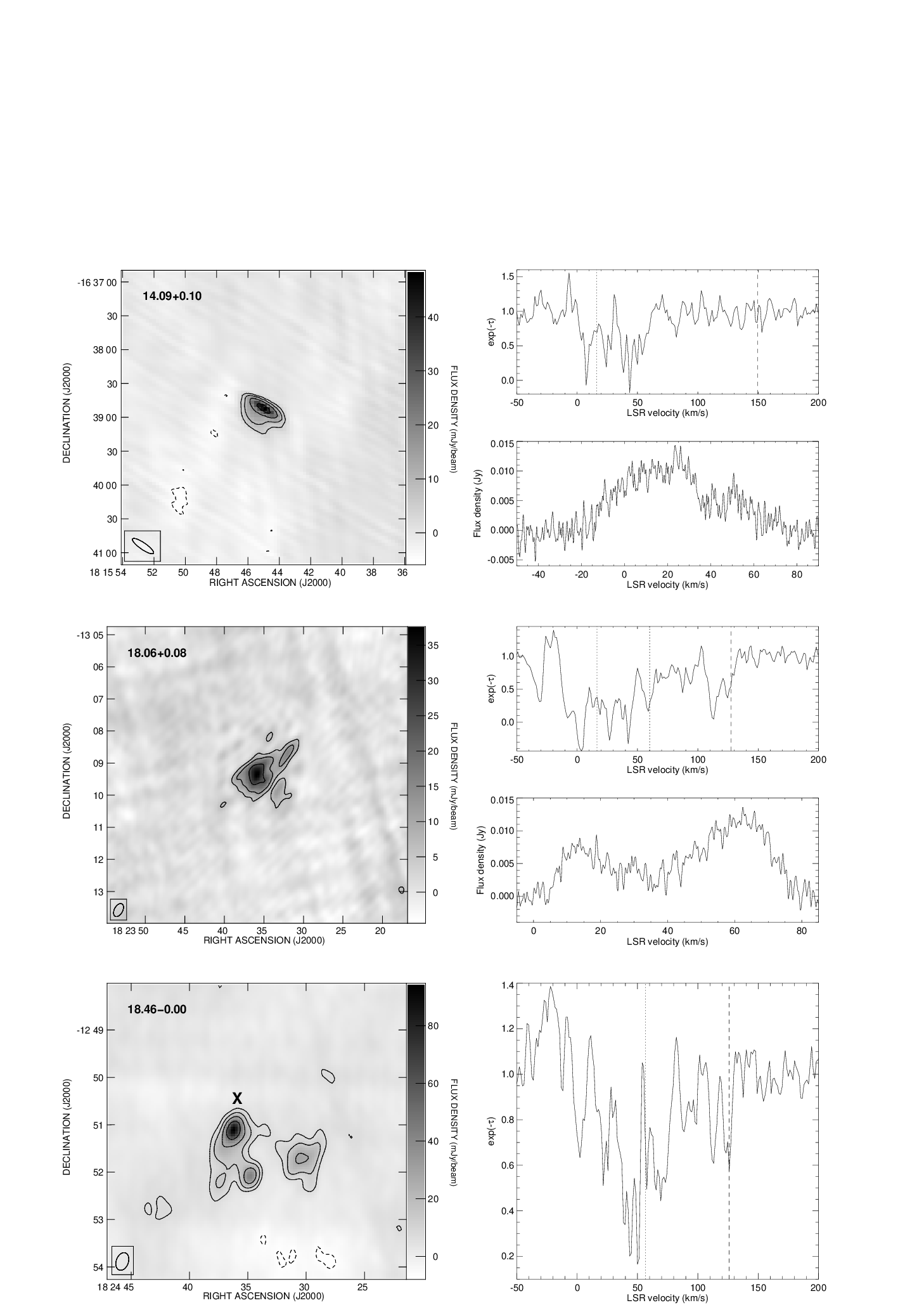}
\caption{Continued.}
\end{figure*}

\setcounter{figure}{0}
\begin{figure*}
\centering
\includegraphics[width=17cm]{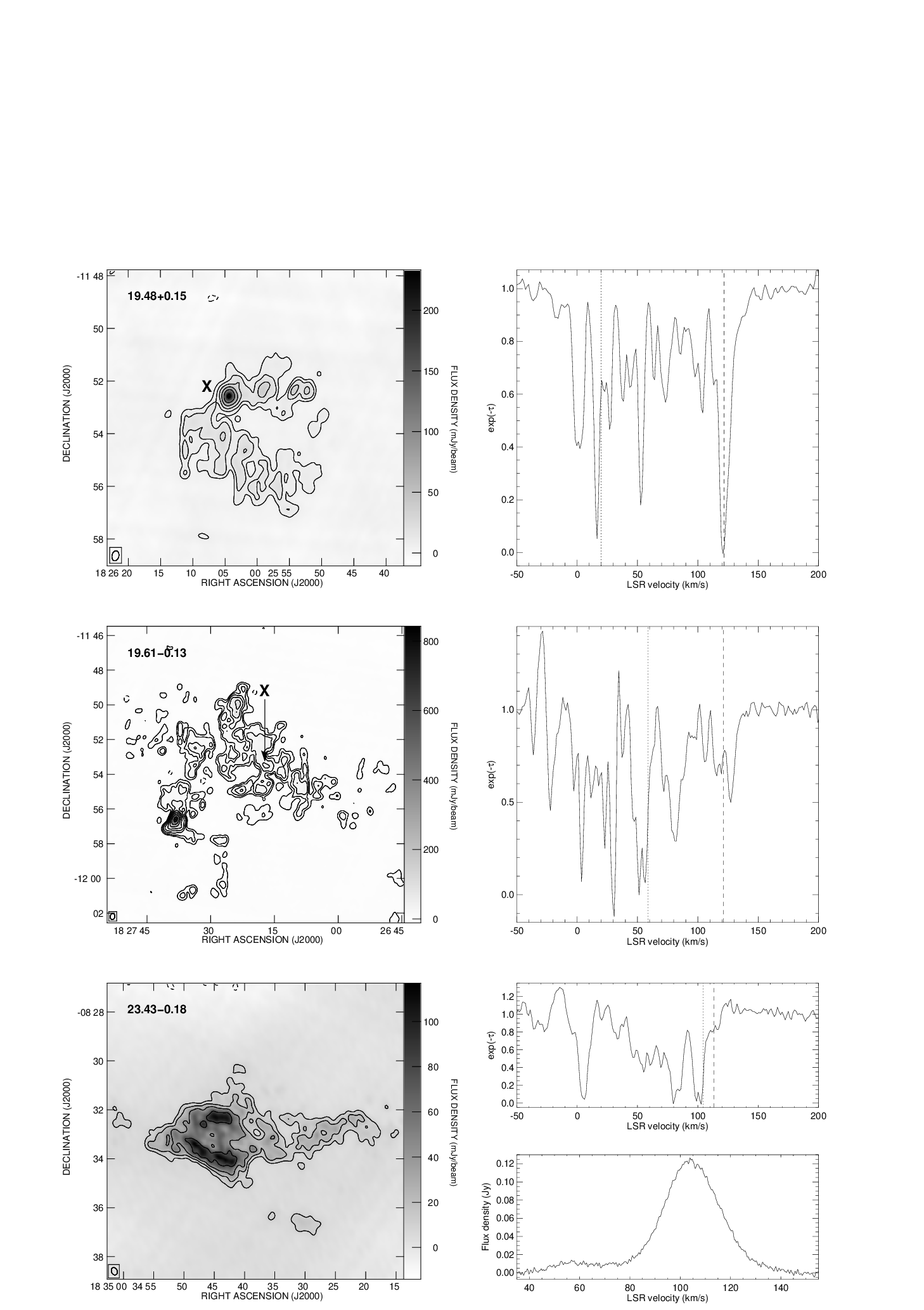}
\caption{Continued.}
\end{figure*}

\setcounter{figure}{0}
\begin{figure*}
\centering
\includegraphics[width=17cm]{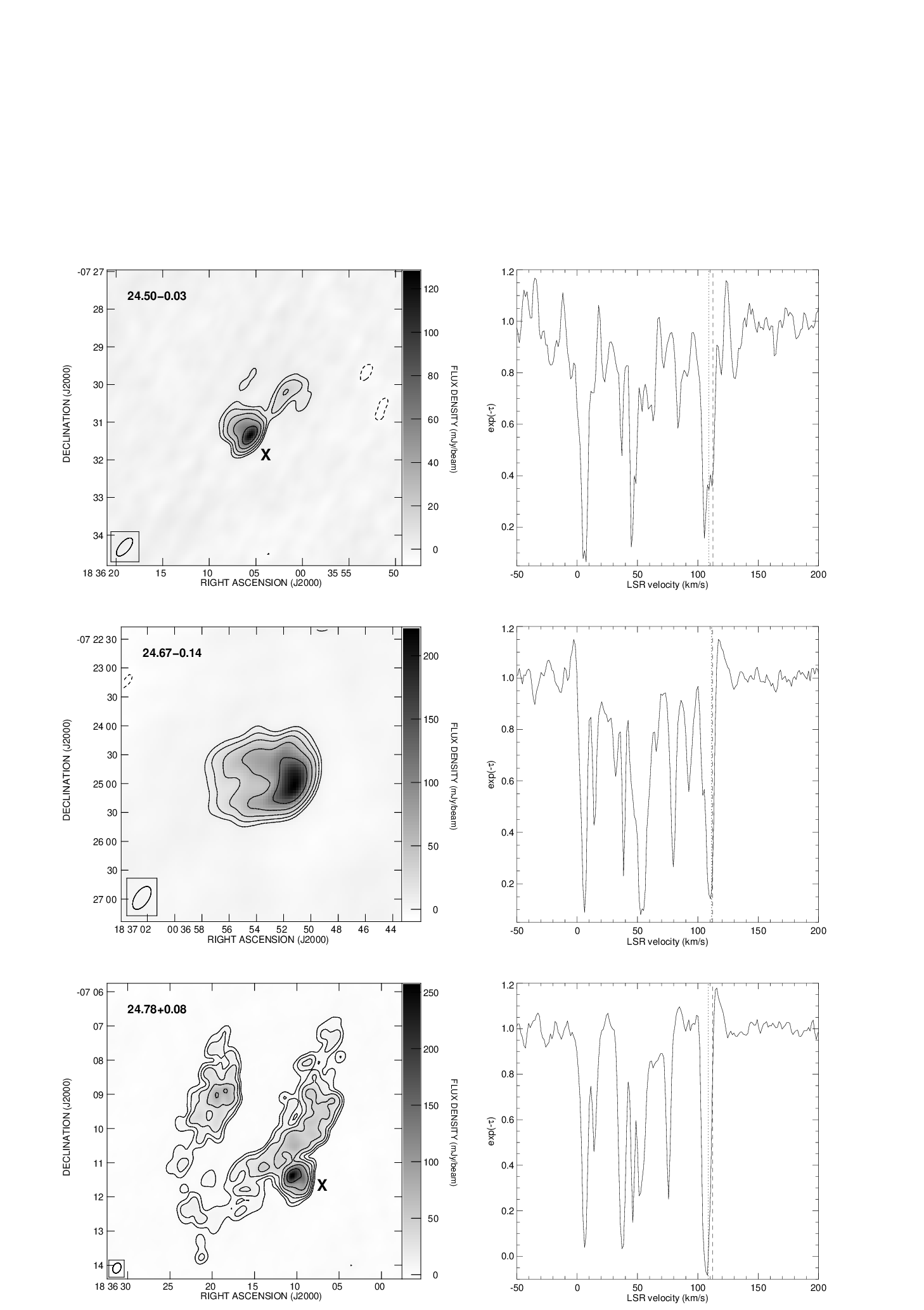}
\caption{Continued.}
\end{figure*}

\setcounter{figure}{0}
\begin{figure*}
\centering
\includegraphics[width=17cm]{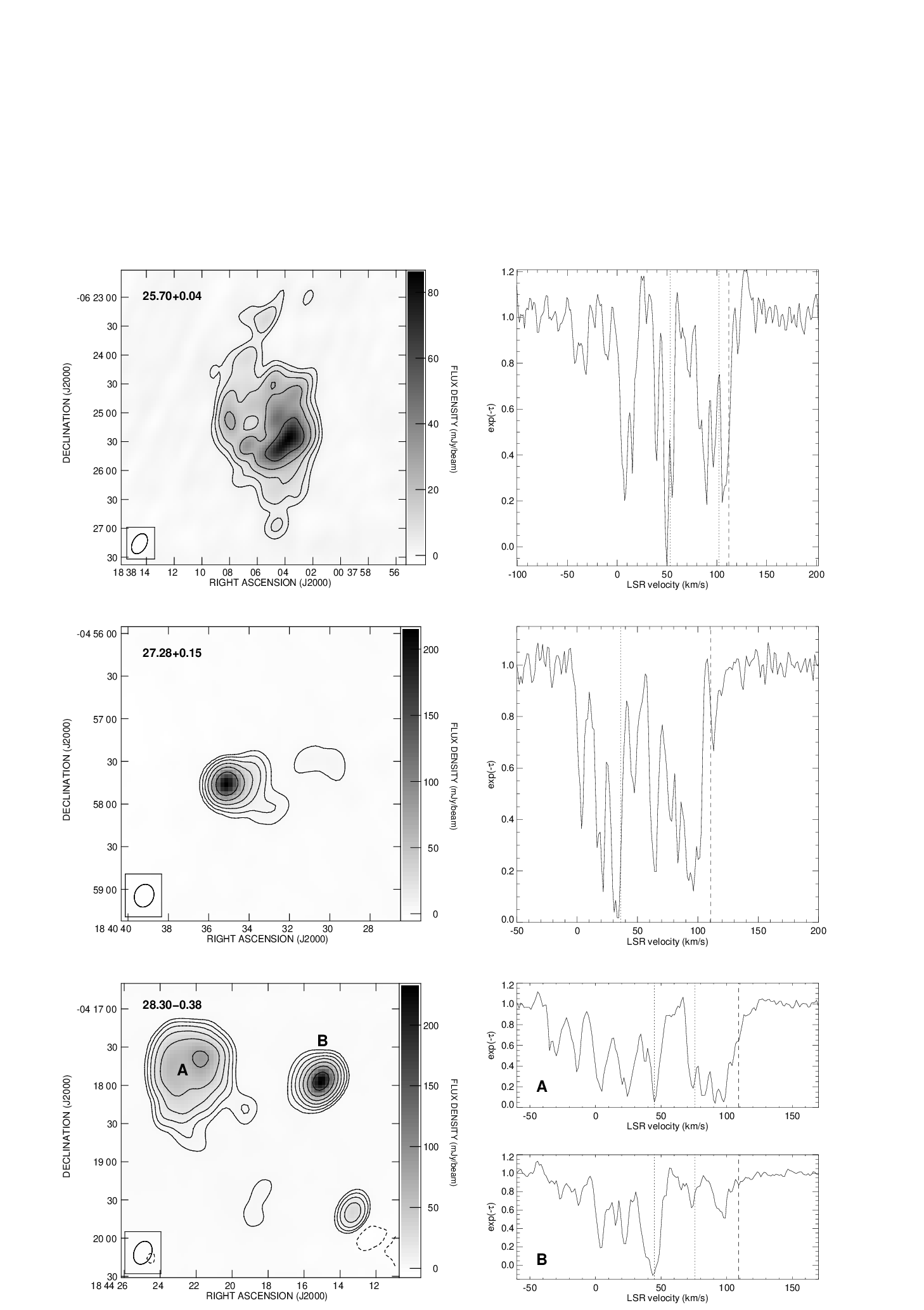}
\caption{Continued.}
\end{figure*}

\setcounter{figure}{0}
\begin{figure*}
\centering
\includegraphics[width=17cm]{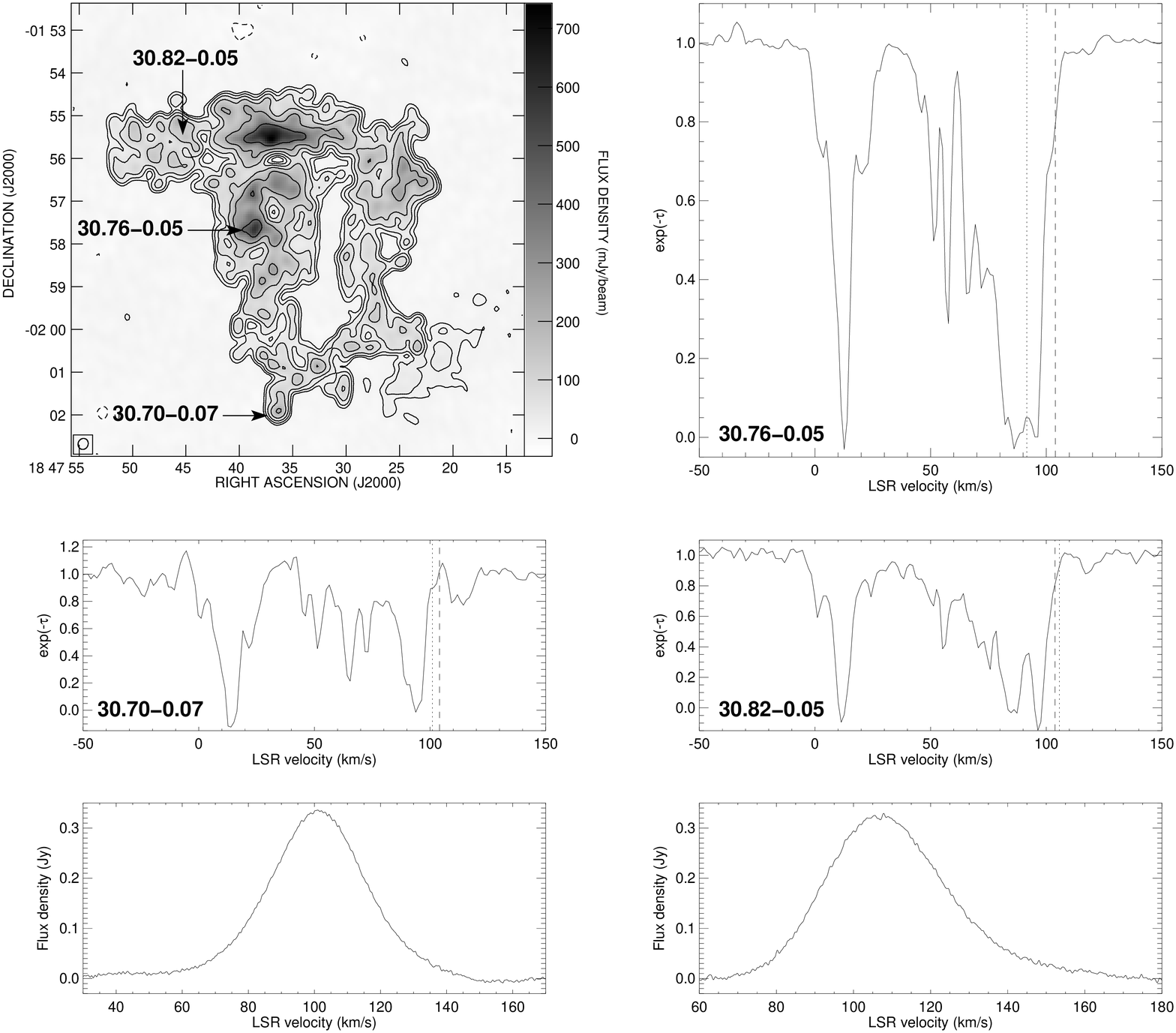}
\caption{Continued.}
\end{figure*}

\setcounter{figure}{0}
\begin{figure*}
\centering
\includegraphics[width=17cm]{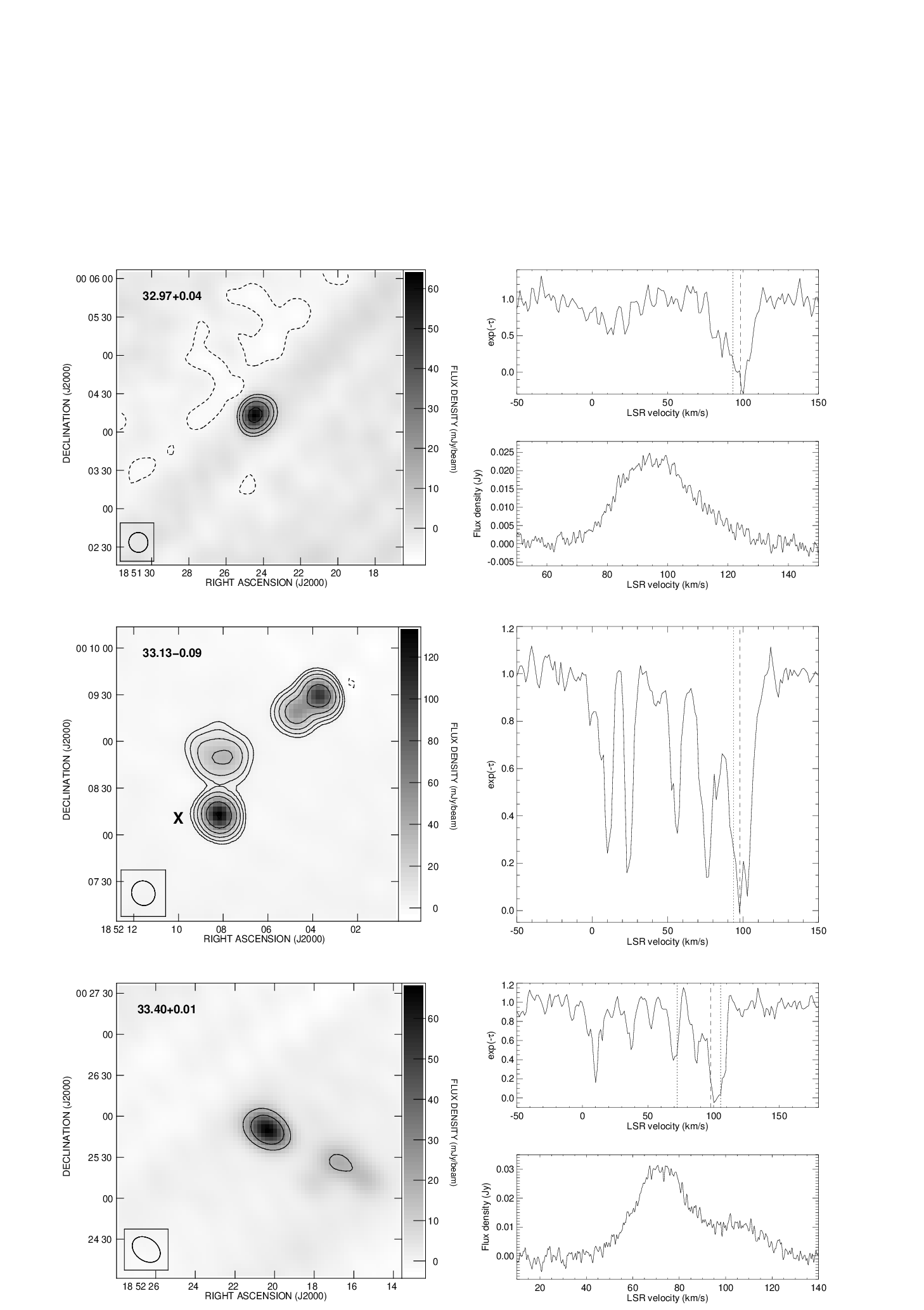}
\caption{Continued.}
\end{figure*}

\setcounter{figure}{0}
\begin{figure*}
\centering
\includegraphics[width=17cm]{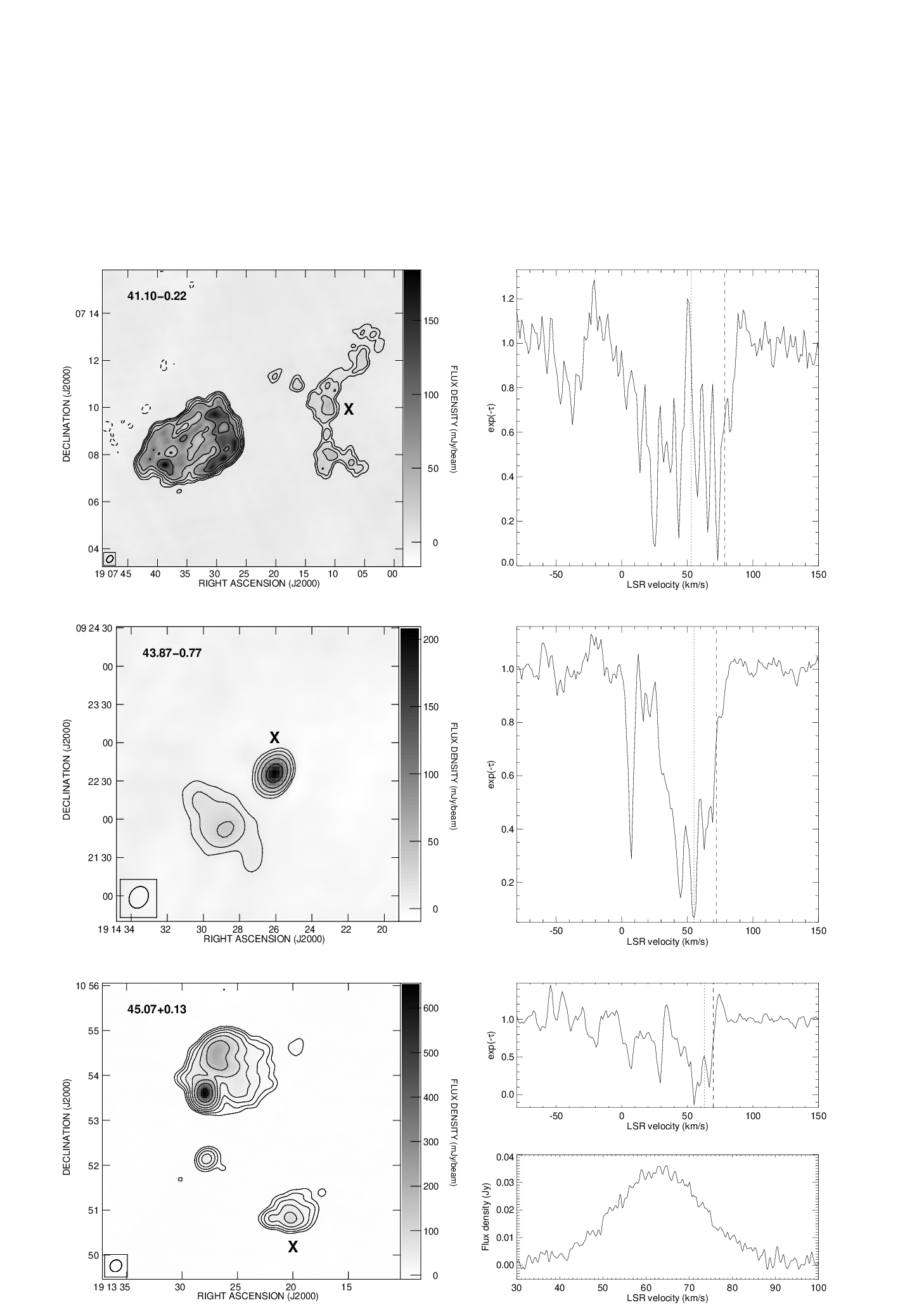}
\caption{Continued.}
\end{figure*}

\setcounter{figure}{0}
\begin{figure*}
\centering
\includegraphics[width=17cm]{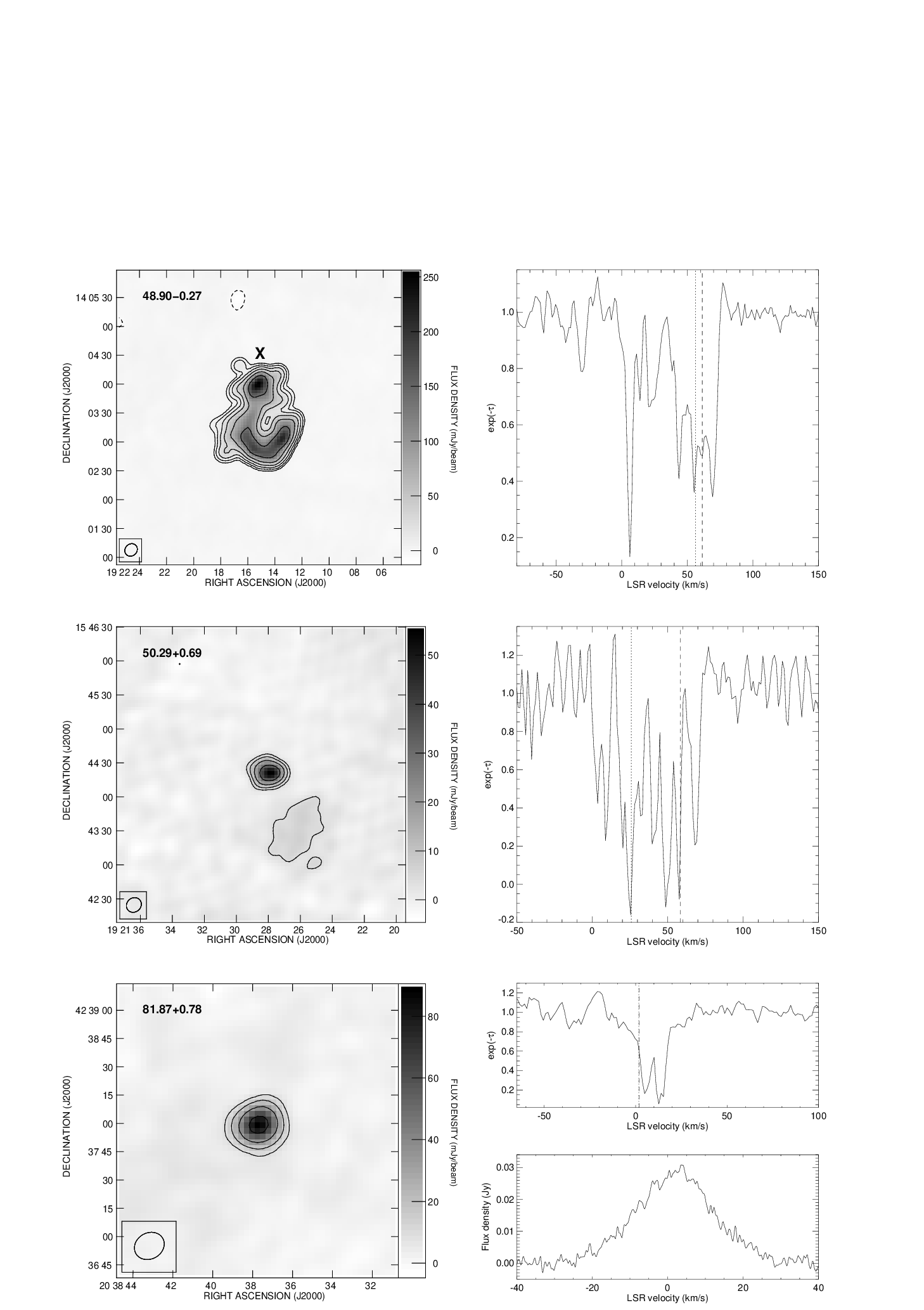}
\caption{Continued.}
\end{figure*}

\subsection{Notes on Selected Sources}

{\it 10.32--0.15 and 10.62--0.38} -- The methanol maser emission in 10.32--0.15 and 10.62--0.38 range from 4 to 17 \kms~and --10 to 7 \kms~respectively. The recombination line velocities from L89 for the two sources are 11.0 and 0.4 \kms~respectively. Both sources are associated with the W31 complex, the distance to which has been a matter of contention. Absorption line experiments have not discovered any absorption features beyond $\sim 45$ \kms. A purely straightforward resolution of the KDA would put both sources at the far distance. However, it is intriguing that no \ion{H}{i} absorption is seen over a velocity interval corresponding to $\sim 7$ kpc, leading to suggestions that the source is at the near distance with the systemic velocity differing from circular motion by $\sim 35$ \kms~(e.g. \citealt{wils74}). \citet{corb04} suggest that the lack of absorption features in both atomic and molecular gas is due to a large gas hole between Galactic longitudes of 5\degr and 25\degr~at velocities greater than 45 \kms. However, there are a number of sources in the same Galactic longitude range that {\it do} show \ion{H}{i} absorption features between 45 \kms~and the tangent point velocity (as can be seen from Fig. 1).  This casts doubt on this explanation, although it is possible that there is a lack of \ion{H}{i} in the specific direction of W31. 

The situation is complicated by the observation of the nearby luminous blue variable star LBV 1806-020 to be at the far distance of $15.1^{+1.8}_{-1.3}$ kpc (\citealt{corb04}; as inferred by the presence of ammonia absorption at 73 \kms) and the derivation of a spectrophotometric distance to the \ion{H}{ii} region G10.2--0.3 to be $3.4 \pm 0.3$ kpc \citep{blum01}, although both sources have similar LSR radial velocities. This suggests a superposition of complexes at vastly different distances which coincidentally have very similar radial velocities. A parallax measurement of the distances to the regions inside this complex is thus required to determine the true distances and the peculiar motions of the various sub-regions. In light of the fact that \ion{H}{i} absorption {\it is} seen at velocities higher than 45 \kms~at similar Galactic longitudes (10.47+0.02 for example), we suggest that both these regions are at the same distance as G10.2--0.3, 3.4 kpc. The difference between the LSR velocity expected for a source at that distance, and the observed LSR velocity can then be attributed to peculiar motions. For example, W3OH has a large peculiar motion of 17 \kms~towards the Galactic center in addition to rotating 14 \kms~slower than the Galactic rotation curve \citep{xu06}. We stress that an independent study that is not based on kinematics is required to resolve the conflicting ideas on the true distance to the regions in the W31 complex.

{\it 12.20--0.10} -- There are two recombination line velocities detected here at 27.0 and 53.6 \kms. The methanol maser emission extends from 2 to 15 \kms~and is thus associated with the 27.0 \kms~\ion{H}{ii} region. The continuum image shows multiple components, which could give rise to the two recombination line systems. An alternate possibility is that one of the \ion{H}{ii} regions giving rise to the recombination lines is optically thick and is undetected at 21 cm. This is possible as the measurements of L89 were made at 3.6 cm, and an optically thick \ion{H}{ii} region can be 30 times weaker at 21 cm compared to 3.6 cm. The \ion{H}{i} spectra from all components of the 21 cm image show absorption features between the systemic velocity and the tangent point placing the source(s) at the far distance. Hence, while it is possible that one of the \ion{H}{ii} regions is not detected in the continuum image at 21 cm, we tentatively place both regions including the methanol maser at the far distance.

{\it 12.68--0.18} -- The GBT spectrum (at 6 cm) shows two recombination line systems, at 32.0 and 57.4 \kms, the 32.0 \kms~component being weaker. The methanol maser emission extends from 50 to 62 \kms~and is thus associated with the 57.4 \kms~recombination line. The 21 cm continuum image shows a very extended structure to the source, the component marked `X' being closest to the methanol maser position. The \ion{H}{i} spectrum for this source shows absorption features up to $\sim 55$ \kms, which is consistent with it being associated with the 57.4 \kms~region and it being at the near distance. The signal to noise ratio at the other continuum regions is too low to determine whether there is another type of \ion{H}{i} spectrum that is associated with the 32.0 \kms~component. Hence, we classify the methanol maser to be at the near distance, and do not make any claim about the second \ion{H}{ii} region in the line of sight.

{\it 18.06+0.08} -- The GBT spectrum shows two velocity components, at 16.4 and 60.1 \kms. The methanol maser, whose emission ranges from 45 to 57 \kms, is associated with the 60.1 \kms~component. The 21 cm continuum is dominated by a single source, although it is possible that some of the structure (especially the extension to the north-west) arises from a superposition of a second source. In the absence of a 6 cm image, we are unable to determine whether both \ion{H}{ii} regions are seen at 21 cm. Only the central source has adequate signal to noise ratio to resolve the KDA, the spectrum from which is shown in Fig. 1. The spectrum would place the source at the far distance for either RRL velocity, and thus we indicate the two possibilities in Table 1 without making any claim regarding which component is seen in our image.

{\it 23.43--0.18} -- The GBT spectrum shows a prominent component at 104.5 \kms~which is associated with the maser emission which spans a velocity range between 94 to 113 \kms, along with a much weaker component at 60.6 \kms. The tangent point velocity is 113.6 \kms. This source is classified at the near distance by \citet{sewi04} and \citet{kolp03}. However, our absorption spectrum shows an absorption component at 116.5 \kms. We thus assign the far distance with quality grade B to this source.

{\it 24.50--0.03} -- This source, with a systemic velocity of 109.2 \kms~(maser emission ranges from 108 to 116 \kms) is classified at the near distance by \citet{sewi04}. The tangent point velocity at this Galactic longitude is 112.9 \kms. Our \ion{H}{i} spectrum shows an absorption feature at around 130 \kms detected at greater than 3$\sigma$ level. We thus classify this source to be at the far distance with quality grade A.

{\it 24.67--0.14} -- The systemic velocity of this source is 111.4 \kms, while the tangent point is at 112.9 \kms. The highest velocity of an absorption feature is 110.5 \kms, and thus this source is classified at the near distance with quality grade B.

{\it 24.78+0.08} -- The recombination line velocity is 108.6 \kms, and \ion{H}{i} absorption is seen up to 108.0 \kms. Since the tangent point velocity is 112.8 \kms~, we classify this source at the near distance with quality grade B.

{\it 25.70+0.04} -- There are two recombination line systems at this position, at 53.3 \kms~and 102.0 \kms. The methanol maser emission ranges from 89 to 101 \kms, and is thus associated with the 102.0 \kms~recombination line. The 21 cm continuum image shows a single source with some sub-structure. \ion{H}{i} absorption at greater than 3$\sigma$ level is seen up to 120.9 \kms, while the tangent point velocity is 112.4 \kms. This would place either recombination line source at the far distance with quality grade A. \citet{down80} list a single H$110\alpha$ velocity of 52.0 \kms~which suggests that the region seen at 21 cm is not associated with the methanol maser. Thus, we do not give a distance to the methanol maser, and classify the 53.3 \kms~component to be at the far distance.

{\it 28.30--0.38} -- L89 report two recombination lines at 44.8 \kms~and 75.6 \kms. The methanol maser emission, ranging from 79 to 93 \kms~is thus associated with the 75.6 \kms~component. There are two distinct continuum sources (labeled A and B in Figure 1) seen in the 21 cm image, and both display \ion{H}{i} spectra that place the source at the far distance. Hence, while we are unable to determine which continuum source gives rise to each recombination line, we classify both sources at the far distance.

{\it 30.70--0.07, 30.76--0.05 and 30.82--0.05} -- The three regions are part of W43, a large and complex region as seen in Figure 1. Absorption features are seen at velocities greater than the tangent point velocity of 104.5 \kms~for 30.70--0.07 and 30.82--0.05, while for 30.76--0.05, absorption is seen up to 101.5 \kms. The systemic velocities for 30.70--0.07, 30.76--0.05 and 30.82--0.05 are 101.0, 91.6 and 105.7 \kms~respectively. A straightforward application of the KDA resolution would put 30.70--0.07 and 30.76--0.05 the far distance (with quality grades A and B respectively), and 30.82--0.05 at the tangent point. A stronger source in the same complex, 30.78--0.03 is also found to be at the far distance (with a systemic velocity of 91.6 \kms) by \citet{kolp03}. Since the regions are part of the same complex, their different velocities can be explained as arising from the random motions within the complex. Hence, we assign all sources at the far distance corresponding to 97.5 \kms~which is the mean velocity of the different components.

{\it 32.97+0.04} -- The recombination line velocity measured at GBT is at 93.2 \kms, while the tangent point velocity is 99.0 \kms. \citet{wats03} do not classify the nearby (and presumably the same) source 32.99+0.04 since its systemic velocity is within 10 \kms~of the tangent point velocity. \ion{H}{i} absorption can be seen up to at least 104.1 \kms. Thus, we classify this source at the far distance with quality grade B.

{\it 33.13--0.09} -- This source is classified as being at the tangent point by \citet{aray02} and \citet{kuch94}. The L89 recombination line velocity is at 93.8 \kms, and \ion{H}{i} absorption is seen up to 102.8 \kms. Since the tangent point velocity is 98.6 \kms, we classify this source at the far distance with quality grade B.

{\it 33.40+0.01} -- The GBT spectrum shows a prominent line at 72.3 \kms, and a weaker component at 105.4 \kms. The methanol maser, whose emission ranges from 96 to 107 \kms~is associated with the latter component. The 21 cm continuum shows a compact source with a much weaker component to the southwest, the latter being too weak to extract a \ion{H}{i} spectrum. The tangent point velocity at this Galactic longitude is 97.9 \kms, which would place the methanol maser at the tangent point. \citet{kuch94} list the nearby (and potentially the same) source G33.4-0.0 (with Galactic latitude and longitude of 33.42 and 0.0 respectively) to be at a velocity of 76.5 \kms~and being at the far distance. It is thus possible that the brighter continuum source is responsible for the 72.3 \kms~recombination line, and the weaker recombination line arises from the weak continuum source or is undetected at 21 cm. Hence, we classify the 72.3 \kms~component at the far distance, and locate the methanol maser at the tangent point.

{\it 48.90--0.27} -- This source, at a systemic velocity of 66.5 \kms, is part of the W51 complex. Kinematically, this source would be at the tangent point since the tangent point velocity is 62.2 \kms~(and has been classified so by \citealt{kolp03} and \citealt{kuch94}). However, \citet{genz81} measured a distance of $7 \pm 1.5$ kpc using statistical parallax by VLBI measurements of H$_2$O masers in W51 MAIN. Since this is a measurement that is independent of the Galactic rotation curve or other distance indicators, we adopt this as the distance to this source.

{\it 50.29+0.69} -- The H110$\alpha$ recombination line measured by \citet{aray02} is at 25.9 \kms, while the methanol maser emission extends from 26 to 33 \kms. This source is classified at the near distance by \citet{wats03} and \citet{aray02} using formaldehyde absorption. The \ion{H}{i} absorption spectrum however clearly places this source at the far distance with quality grade A. This is an example (albeit rare) of how the lack of formaldehyde clouds between the near and far distance points can give rise to incorrect classification of a source.

{\it 81.87+0.78} -- The methanol maser emission in this source ranges from 0 to 13 \kms~while the recombination line velocity is 1.7 \kms. The \ion{H}{i} spectrum shows absorption up to 15.2 \kms, and the tangent point velocity is 3.0 \kms. Hence, we classify this source at the far distance with quality grade B. The use of the modified technique to calculate kinematic distances, described in this section 4, would however place this source at the tangent point, as is noted in Table 1.

\section{Discussion}

16 out of 38 sources in our study with KDA resolutions have been targeted in previous studies. Sources for which our classification disagrees with that of previous work are noted in section 5.1. Figure 2 shows the locations of the sources in a face-on diagram of the Galaxy. We also compiled a list of \ion{H}{ii} regions whose KDA has been resolved by \citet{kuch94}, \citet{aray02}, \citet{fish03}, \citet{kolp03}, \citet{wats03}, and \citet{sewi04} in addition to those in our study. All the distances were recalculated using the C85 rotation curve, and the modified technique described in section 4. Some KDA resolutions, especially of sources with velocities near the tangent point velocity, were reinterpreted by the criteria described in section 4 for consistency. Figure 3 shows all the sources in this compilation (we do not distinguish the sources by author since a number of sources are shared by multiple groups) superposed with two popular spiral arm models of the Galaxy -- the NE2001 model of \citet{cord02} and the model of \citet{vall02,vall95}. Figure 4 shows the same sources, but with distances calculated using the modified procedure. The NE2001 model defines spiral arms as logarithmic spirals with perturbations similar to those described in \citet{tayl93}. The \citet{vall02,vall95} model defines the spiral arms as logarithmic spirals, the parameters of which are determined from a meta-analysis of numerous observational datasets. It is to be noted that \citet{vall02,vall95} use an $R_0$ of 7.2 kpc, and consequently, we have scaled the model to $R_0$ of 8.5 kpc for comparison with our data and the NE2001 model.

The data points in Figs. 3 and 4 sample the Crux-Scutum, Carina-Sagittarius and Perseus arms. Qualitatively, the NE2001 model is a better description of the Galactic spiral structure compared to the model of \citet{vall02}. The discrepancy between the two models is greatest for the Carina-Sagittarius arm. A number of \ion{H}{ii} regions also appear to lie in inter-arm regions, although the error bars in the distances are almost always large enough for these regions to be potentially associated with one or two arms. The lack of data points along the Outer arm could be a sensitivity effect, as only the brightest \ion{H}{ii} region complexes would have radio continuum high enough to be targeted in these studies.

\begin{figure}
\resizebox{\hsize}{!}{\includegraphics{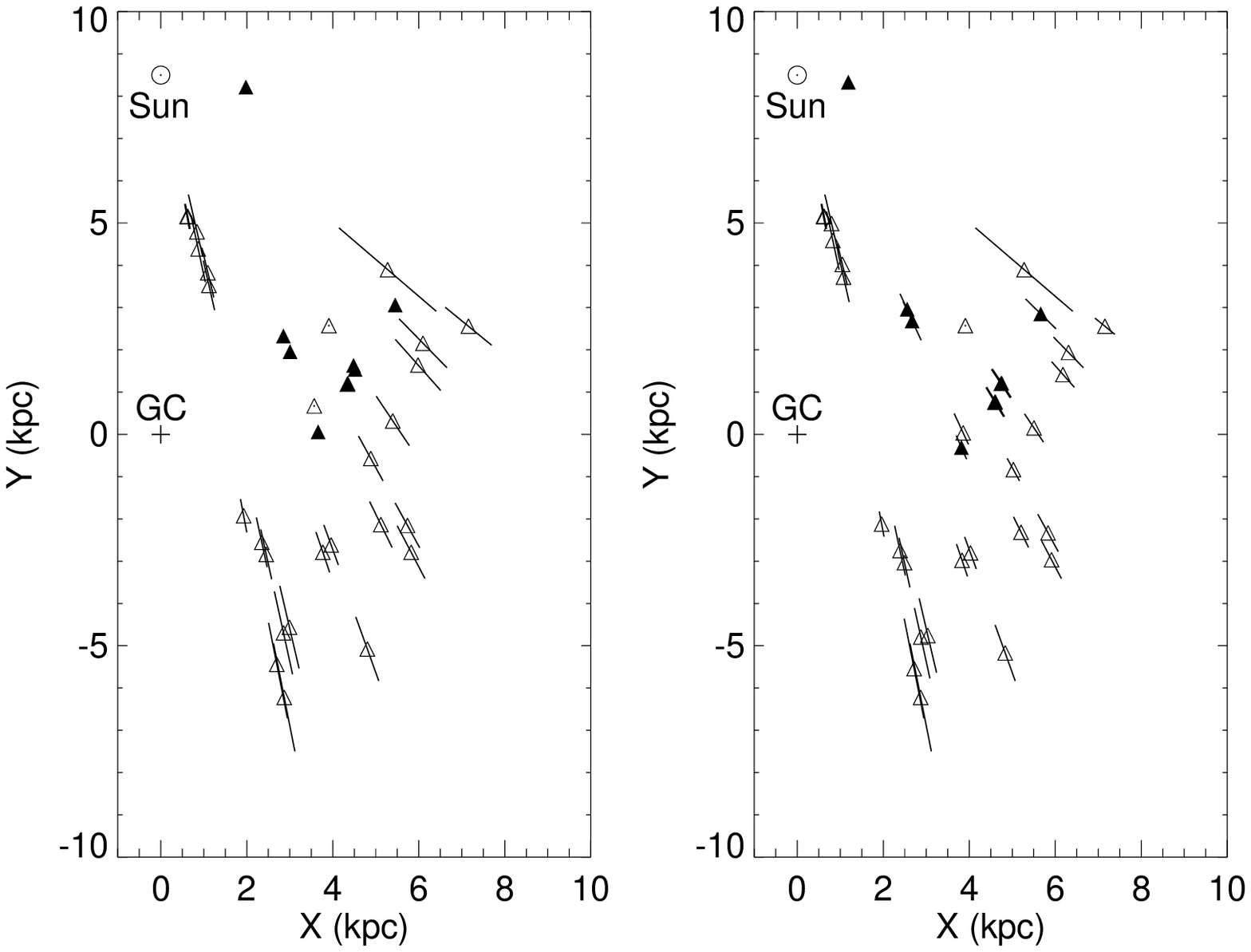}}
\caption{Positions of the 6.7 GHz methanol masers and associated \ion{H}{ii} regions for which we have resolved the KDA. The left panel shows the distances using the conventional method labeled $d$ in Table 1, while the right panel shows the distances calculated using the modified LSR velocity and rotation curve, labeled as $d'$ in Table 1. The open and filled triangles show the distances with quality grades A and B respectively. The location of the Galactic Center is marked as ``GC''.}\label{fig2}
\end{figure}

\begin{figure*}[h]
\centering
\includegraphics[width=17cm]{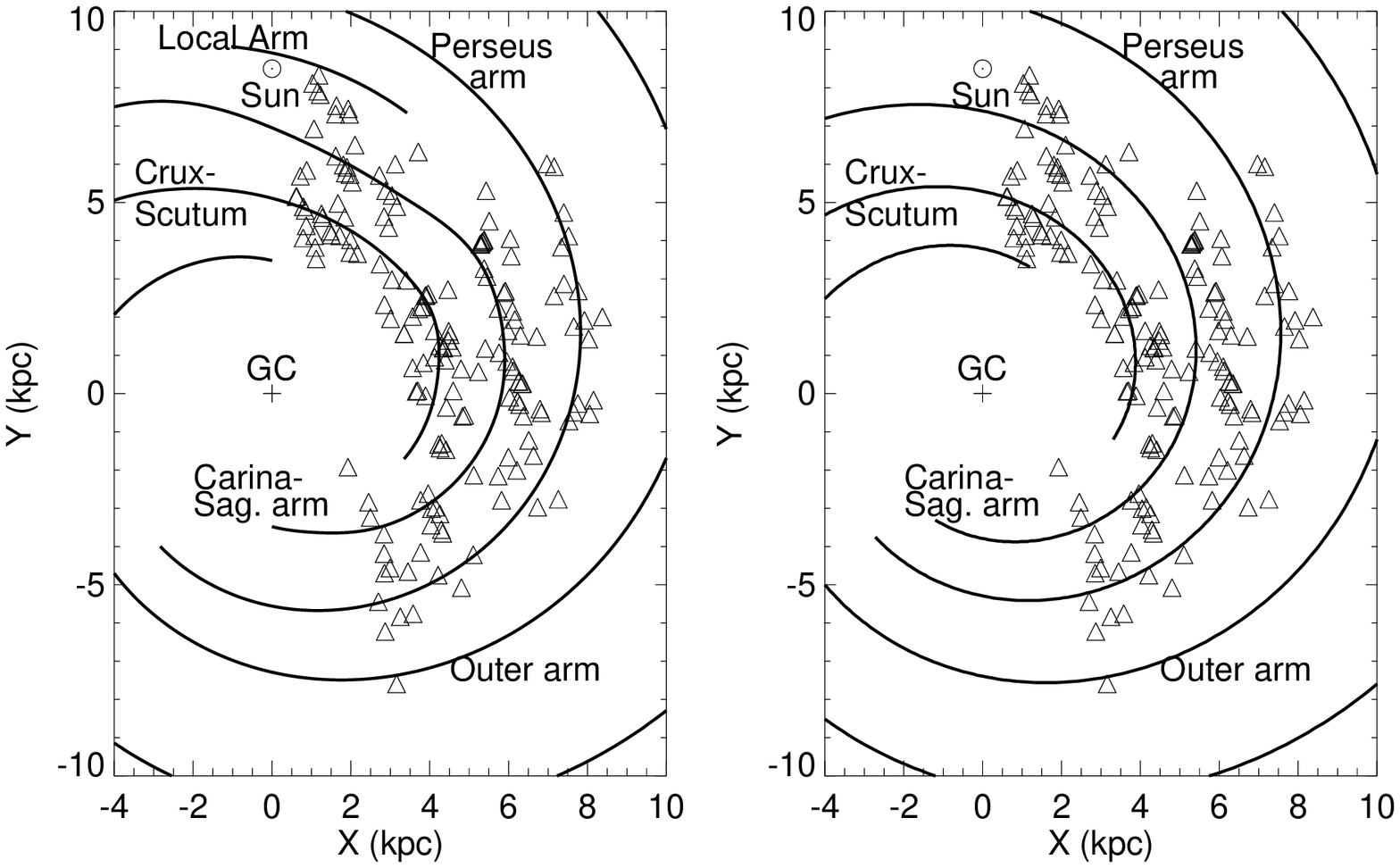}
\caption{\ion{H}{ii} regions with resolved KDA compiled from \citet{kuch94}, \citet{aray02}, \citet{fish03}, \citet{kolp03}, \citet{wats03}, \citet{sewi04} and our study. Distances have been calculated using the C85 rotation curve. The left and right panels show the spiral arm loci of the NE2001 model of \citet{cord02}, and the \citet{vall95} model respectively. For clarity, no error bars are shown.}\label{fig3}
\end{figure*}

\begin{figure*}[h]
\centering
\includegraphics[width=17cm]{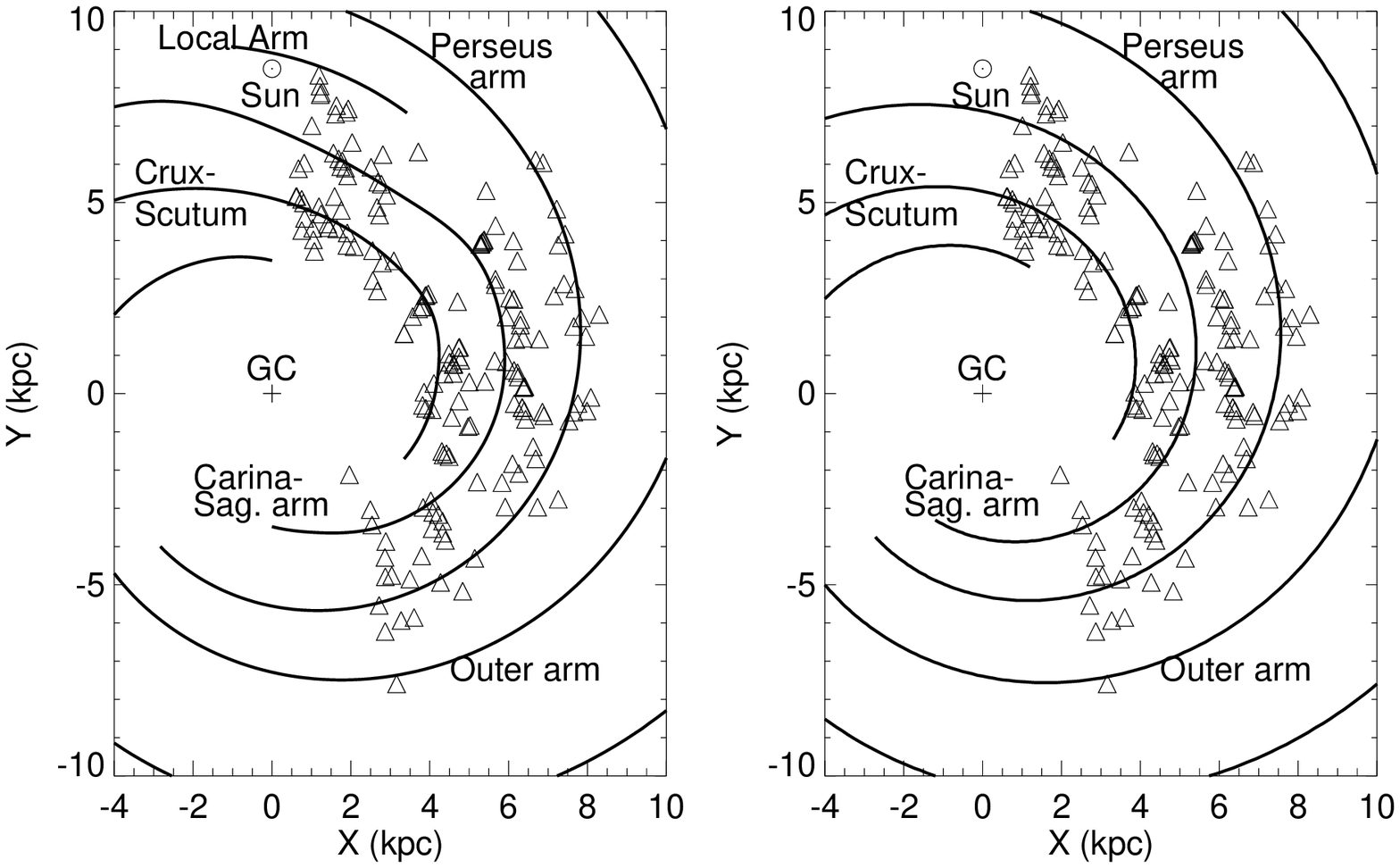}
\caption{\ion{H}{ii} regions with resolved KDA compiled from \citet{kuch94}, \citet{aray02}, \citet{fish03}, \citet{kolp03}, \citet{wats03}, \citet{sewi04} and our study. Modified kinematic distances have been calculated using the procedure described in Section 4. The left and right panels show the spiral arm loci of the NE2001 model of \citet{cord02}, and the \citet{vall95} model respectively.  For clarity, no error bars are shown.}\label{fig4}
\end{figure*}

\citet{fish03} proposed using the small vertical height of compact \ion{H}{ii} regions to resolve the distance ambiguity using a simple relation. Based on a Monte Carlo simulation, they suggest that \ion{H}{ii} regions can be classified to be at the near or far distance based on which distance was closer to a predicted value, $d_p = 1.84~|b|^{-1}$. For our sample, this method has an accuracy of 70\%. Surprisingly, the accuracy drops to 60\% when restricted to sources that have $\Delta d = d_{far} - d_{near} > 4.7$ kpc. In contrast, \citet{fish03} found the method accurate to better than 90\% for a sample with similar criteria in their simulation. It is to be cautioned that our sample has numerous biases, so that it cannot be used to test the validity of the method, although we note that \citet{wats03} had a similar success rate with their sample. Nevertheless, it indicates the danger of using this technique for KDA resolution of specific sources as opposed to a statistical study of a large sample. As such, this specific line of argument used by \citet{corb04} to justify their placing 10.32--0.15 at the far distance and 10.62--0.38 at the near distance is invalid.

It is also curious that we find about 70\% of the sources to be at the far distance. This is somewhat similar to the observation of \citet{kolp03}. However, our methanol maser sample is not flux limited, and has a numerous selection effects and biases associated with it. Hence, we do not attempt to draw any conclusion from this observation.

\subsection{Application to Fainter Sources}
As mentioned earlier, the sources selected for inclusion in our study involved methanol masers in the first Galactic quadrant and at $l> 10\degr$, with direct or nearby associations of NVSS sources that were stronger than 30 mJy. This resulted in a selection of fewer than 30\% of the masers in this region from the General Catalog, of which about 40\% were too resolved in our data to solve the KDA. To extend this study to a larger sample of methanol masers, one would need to observe fainter sources. This raises the question of plausibility of applying the technique to faint continuum sources.

One faces two challenges when dealing with faint sources. Firs, there is significant contamination from the stronger sources (which are numerous in the Galactic Plane) through sidelobes of the dirty beam. Second, there is low-level \ion{H}{i} emission in the entire image. As explained in section 3, CLEANing the data is essential to be able to remove the contamination from stronger sources. However, a full field CLEAN does not work well for faint sources that are significantly contaminated by stronger sources in the field. On the other hand, a CLEAN with boxes at the locations of the continuum emission will tend to bring \ion{H}{i} emission, which has a distribution that is completely different from that of the continuum, into the CLEAN boxes and into the target source.

The problem of emission can be mitigated to a certain extent by going to a larger array configuration, which tends to filter out more of the extended Galactic emission. Doing this would, however, resolve the sources even further, requiring significantly greater telescope time to achieve sufficient signal to noise. This is especially so, since a $3\sigma$ detection of absorption at optical depths corresponding to $\exp(-\tau)$ of at least 0.3 is required to resolve the KDA with high confidence. A lower configuration (D or the proposed E configuration) would yield higher signal to noise ratio on the sources by keeping them unresolved, at the expense of introducing a larger extent of \ion{H}{i} emission.

Extending \ion{H}{i} absorption against radio continuum emission for resolution of the KDA is thus a significant challenge in the case of weak continuum emission. A promising variant is to employ \ion{H}{i} self-absorption to distinguish between near and far distances, a technique which has recently been used to resolve the KDA towards MYSOs by \citet{busf06}. Since this method relies on the absorption (or the lack thereof) of background \ion{H}{i} line emission at the systemic velocity of the source, it does not suffer from the problem of lack of 21 cm continuum. The only challenge is in the interpretation of the \ion{H}{i} features and distinguishing self-absorption from the often complex emission line profile. \citet{busf06} estimate a success rate of $\sim 70$\% for KDA resolution using this technique for their sample of MYSOs selected from the red MSX source (RMS) survey.  Formaldehyde absorption at 6 cm may also be viable for sources having weak 21 cm emission. An advantage is that young \ion{H}{ii} regions, whose flux density has a $\sim \nu^2$ dependence on frequency would be brighter at 6 cm, but a potential problem for the weaker sources is the increased risk of confusion from absorption against the CMB.

\section{Conclusion}
We have resolved the KDA towards 34 6.7 GHz methanol masers and 4 additional \ion{H}{ii} regions using \ion{H}{i} absorption measured with the VLA. Over 70\% of our sources have been observed to be at the far distance. The use of Galactic latitude to resolve the KDA \citep{fish03} is not very successful for our sample. The estimates of distances to the methanol maser sources obtained by resolving the KDA are crucial for the ongoing project to determine the methanol maser luminosity function.

\begin{acknowledgements}
We would like to thank Dana Balser and Ron Maddalena for help with the GBT observations and data reduction. This work was supported in part by the Jet Propulsion Laboratory, California Institute of Technology, under a contract with the National Aeronautics and Space Administration. This research has made use of NASA's Astrophysics Data System.
\end{acknowledgements}

\bibliographystyle{aa}
\bibliography{9799refs}

\end{document}